\documentclass[%
 aip,
 jcp,
 reprint,
 amsmath,amssymb,
]{revtex4-2}

\usepackage[T1]{fontenc}
\usepackage{graphicx}
\usepackage{dcolumn}
\usepackage{bm}
\usepackage{siunitx}
\usepackage{xcolor}

\begin{document}

\preprint{AIP/123-QED}

\title{A classical trajectory method for diatom-atom-atom ternary association reactions}


\author{Daniel Julian}
\author{Jes\'us P\'erez-R\'ios}
\email{jesus.perezrios@stonybrook.edu}
\affiliation{Department of Physics and Astronomy, Stony Brook University, Stony Brook, New York, 11794-1901, United States}

\date{\today}

\begin{abstract}
We introduce a classical trajectory method for calculating direct diatom-atom-atom ternary association reaction rates that incorporates the rotational degrees of freedom of the diatomic reactant. By treating the diatomic molecule as a rigid rotor, we include the effects of the anisotropy of the molecule-atom interaction and the dynamical effects of the rotational motion. We apply this approach to a system of experimental interest, CD$^+$ + He + He, along with the isotopic variant CH$^+$ + He + He $\rightarrow$ CHHe$^+$ + He, and examine the role of the internal, rotational degrees of freedom of the diatomic molecule. In particular, we find that in the regime where the collision energy is much smaller than the rotational energy of the entrance channel, the rotational motion leads to deviation from the classical threshold law by violating the assumptions of long-range capture theory models. 
\end{abstract}

\maketitle

\section{Introduction}

Ternary association reactions, in which two reactants form a bound complex while a third body carries away the excess kinetic energy, are central to a wide range of physical systems, spanning from the formation of molecular ions in cold environments~\cite{Krukow2016,Weckesser2021,Tomza2019,JPR_2021} to volumetric neutralization processes in plasmas~\cite{Kaganovich2010,Bates_plasma,Eletskii_1996}. Building on the seminal contributions of Troe, Herbst, and Bates~\cite{Troe_1977,Troe_1986,Troe2005,Herbst1980,Bates1979,Bates1979bis}, these reactions have been predominantly analyzed within the Lindemann--Hinshelwood framework~\cite{LINDEMANN196793,Hinshelwood1926OnTT}, which treats termolecular association as the outcome of two coupled bimolecular steps: (i) encounter and formation of an energized intermediate complex by two reactants, followed by (ii) stabilization of this complex via collisional deactivation by a third body or through radiative emission. However, such approaches necessarily introduce assumptions regarding the structure, energetics, and dynamical properties of the intermediate complex. Moreover, they exhibit notable discrepancies when confronted with specific experimental datasets~\cite{Bates_issues,book_termolecular,Xie_Poirier_Gellene_2003,Herbst_1989}, indicating that a universally applicable theoretical description---particularly for ion--molecule association reactions---remains unavailable. 

An alternative is to describe association as a direct three-body collision, A + B + C $\rightarrow$ products, without presupposing a long-lived intermediate~\cite{Smith_1962,Perez-Rios2014}. This approach has provided insight into the stability of a single ion in an ultracold neutral bath and ion recombination in its parent gas~\cite{Perez-Rios2018,JPR2024}. For atomic systems, it has also been applied to sulfur~\cite{Sulfur} and halogen recombination~\cite{Koots2025} and to the formation of van der Waals dimers~\cite{Mirahmadi2021}. More recently, direct-collision descriptions have offered a perspective on ozone formation~\cite{Ozone}, suggesting a useful route for studying barrierless association reactions involving more complex reactants~\cite{Koots_2026}.

Here, we develop a classical trajectory method in hyperspherical coordinates for diatom-atom-atom association reactions forming trimers. Following an earlier classical trajectory approach~\cite{Perez-Rios2014}, we map the translational degrees of freedom into a 6D space described in hyperspherical coordinates, and here we introduce the rotational degrees of freedom, which are also treated classically but initialized using the quantum mechanically allowed values for the diatom's angular momentum. We focus on rotational motion, although the method can be extended to include molecular vibration. We find that the initial rotational state greatly affects the association rate when the collision energy is below the molecule's initial rotational energy, whereas its effect diminishes at higher collision energies. We apply our method to the reactions CD$^+$ + He + He $\rightarrow$ CDHe$^+$ + He and CH$^+$ + He + He $\rightarrow$ CHHe$^+$ + He, and for the former we compare with the available experimental data~\cite{BRUNKEN201767}.


\section{Methodology}
\subsection{Classical Dynamics}
In our approach, we treat the diatomic molecule as a rigid rotor and incorporate its rotational motion by calculating the rotor's classical dynamics under the influence of an anisotropic potential. The center-of-mass of the diatom and the coordinates of the free atoms correspond to the coordinates of the reactants in the three-body recombination. We denote the two nuclei in the diatomic molecule by their masses $m_1$ and $m_2$ and positions $\mathbf{r}_1$ and $\mathbf{r}_2$ and introduce their associated relative and center-of-mass coordinates: 

\begin{equation}
    \mathbf{r} \ = \ r_e \mathbf{\hat r} \  = \ \mathbf{r}_2  -  \mathbf{r}_1, \ 
    \mathbf{r}_{\text{mol}} \ = \ \frac{m_1 \mathbf{r}_1  +  m_2\mathbf{r}_2}{M_{\text{mol}}}.
\end{equation}
 The fixed length of the rotor is set to the equilibrium value $r_e$, and the mass of the diatom is $M_{\text{mol}} = m_1 + m_2$. The motion of the relative coordinate $\mathbf{r}$ is purely rotational in the rigid rotor approximation, and it is described in terms of its angular velocity $\bm{\omega}$ and angular momentum $\mathbf{J}$, both of which are perpendicular to $\mathbf{\hat r}$:
\begin{equation}
    \mathbf{\dot {\hat r}} \ = \ \bm{\omega} \times \mathbf{\hat r}, \ \mathbf{J} \ = \ \mu_{12}r_e^2 \bm{\omega},
\end{equation}
where we have introduced the two-body reduced mass $\mu_{12}  = \frac{m_1m_2}{m_1 + m_2}$.

The Hamiltonian of the system, including the atoms with masses $m_3$ and $m_4$ and momenta $\mathbf{p}_3$ and $\mathbf{p}_4$, is:
\begin{equation}
    H \ = \ \frac{\mathbf{J}^2}{2\mu_{12}r_e^2} \ + \ \frac{\mathbf{p}_{\text{mol}}^2}{2M_{\text{mol}}} \ + \ \frac{\mathbf{p}_3^2}{2m_3} \ + \ \frac{\mathbf{p}_4^2}{2m_4} \ + \ V(\mathbf{\hat r}, \mathbf{r}_{\text{mol}}, \mathbf{r}_3, \mathbf{r}_4)
\end{equation}
where $\mathbf{p}_{\text{mol}}$ is the momentum of the center-of-mass of the diatomic molecule.

\begin{figure}
    \centering
    \includegraphics[width=1.0\linewidth]{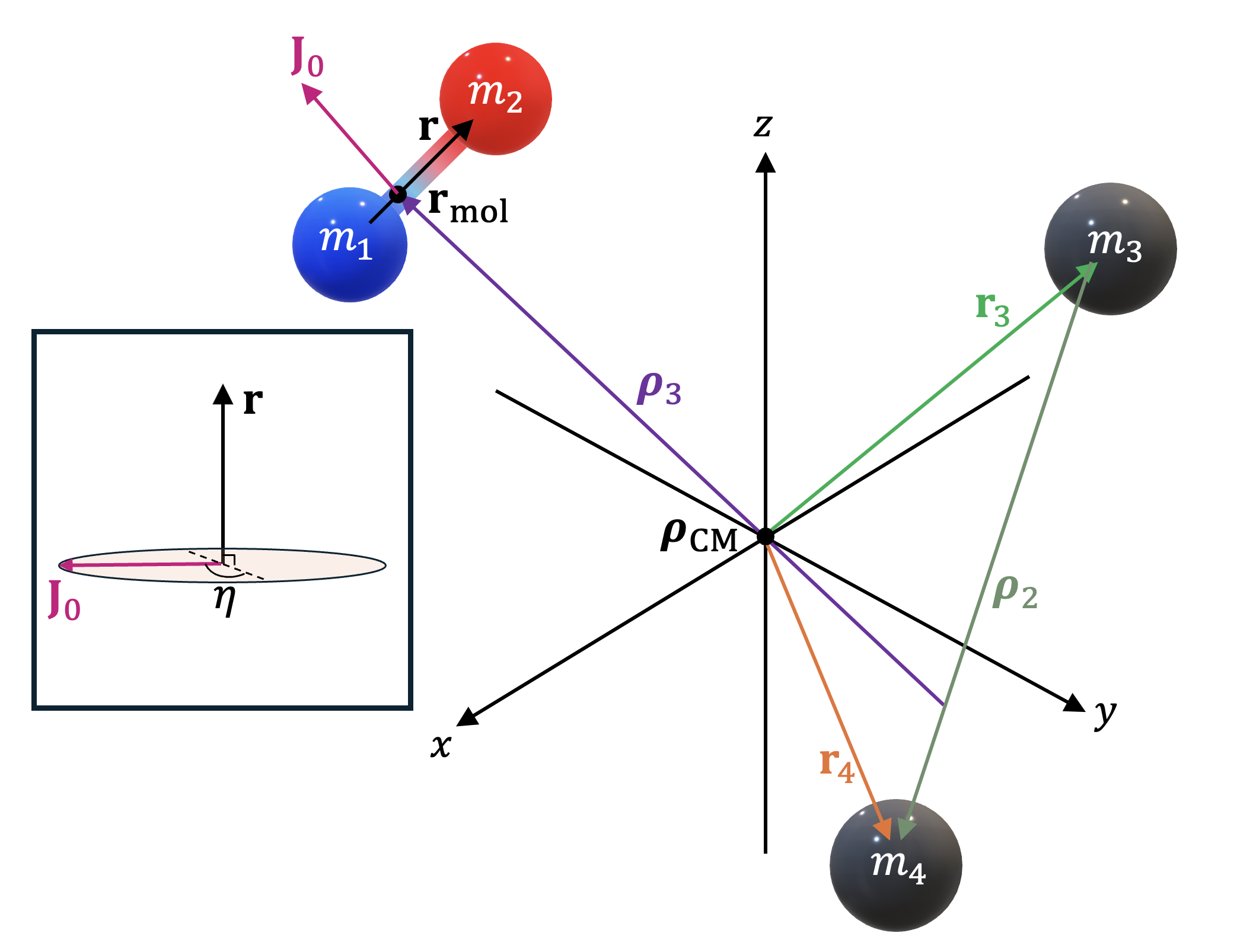}
    \caption{Illustration of the Jacobi coordinates [Eqs.~(\ref{Jacobi})] and the initial diatom angular momentum $\mathbf{J}_0$ in the plane perpendicular to $\mathbf{r}$ with direction given by $\eta$.}
    \label{coords}
\end{figure}

Treating the centers of masses of the reactants as the three bodies, we introduce the Jacobi coordinates for our system (see Fig.~\ref{coords}) as~\cite{Mirahmadi02102022, Py3BR}: 

\begin{equation}
    \begin{split}
    \bm{\rho}_2 \ = \ \mathbf{r}_4 \ - \ \mathbf{r}_3 \\
    \bm{\rho}_3 \ = \ \mathbf{r}_{\text{mol}} \ - \ \frac{m_3 \mathbf{r}_3 \ + \ m_4 \mathbf{r}_4}{m_3 + m_4} \\
    \bm{\rho}_{\text{CM}} \ = \ \frac{M_{\text{mol}}\mathbf{r}_{\text{mol}} + m_3 \mathbf{r}_3 + m_4 \mathbf{r}_4}{M}
    \end{split}
\label{Jacobi}
\end{equation}
Clearly, the Jacobi vector $\bm{\rho}_2$ corresponds to the relative coordinate between the atoms, $\bm{\rho}_3$ is the coordinate for the diatom's center-of-mass relative to the atoms' center-of-mass, and $\bm{\rho}_{\text{CM}}$ is the center-of-mass of the entire molecule-atom-atom system with $M =  M_{\text{mol}}  +  m_3 +  m_4$. In our model, no external field acts on the system and the interactions do not depend on $\bm{\rho}_{\text{CM}}$, so the total linear momentum is conserved and we can take the CM frame and set the origin at the center-of-mass ($\bm{\dot \rho}_{\text{CM}} = \textbf{0}$, $\bm{\rho}_{\text{CM}} = \textbf{0}$). Introducing the reduced masses $\mu_{34}  =  \frac{m_3m_4}{m_3 + m_4}$ and $\mu_{34,\text{mol}}  =   \frac{(m_3+m_4)M_{\text{mol}}}{M}$, we arrive at the Hamiltonian: 
\begin{equation}
    H \ = \ \frac{\mathbf{J}^2}{2\mu_{12}r_e^2} \ + \ \frac{\mathbf{P}_2^2}{2\mu_{34}} \ + \ \frac{\mathbf{P}_3^2}{2\mu_{34,\text{mol}}} \ + \ V(\mathbf{\hat r}, \bm{\rho}_2, \bm{\rho}_3)
\end{equation}
where we have introduced the momenta conjugate to the Jacobi coordinates: $\mathbf{P}_2  =  \mu_{34} \bm{\dot \rho}_2$ and $\mathbf{P}_3  =  \mu_{34,\text{mol}}\bm{\dot \rho}_3$. 

From this Hamiltonian, we obtain the equations of motion for this system:

\begin{equation}
    \begin{split}
    \bm{\dot \rho}_2 \ = \ \frac{\mathbf{P}_2}{\mu_{34}}, \  \mathbf{\dot P}_2 \ = \ -\frac{\partial V}{\partial \bm{\rho}_2} \\ \bm{\dot \rho}_3 \ = \ \frac{\mathbf{P}_3}{\mu_{34,\text{mol}}}, \  \mathbf{\dot P}_3 \ = \ -\frac{\partial V}{\partial \bm{\rho}_3} \\ \mathbf{\dot {\hat r}} \ = \ \frac{1}{\mu_{12}r_e^2} \mathbf{J} \times \mathbf{\hat r}, \  \mathbf{\dot J} \ = \ -\mathbf{\hat r} \times \frac{\partial V}{\partial \mathbf{\hat r}}
    \end{split}
\end{equation}
which are numerically integrated. For the results of this work, RK45~\cite{DORMAND198019} was used for numerical integration with absolute and relative tolerances of 10$^{-12}$ and 10$^{-10}$, respectively. With no explicit time dependence in the Hamiltonian, energy is conserved, and this is checked at the end of each trajectory.

Whereas the equations of motion make use of the Jacobi coordinates, it is preferable to sample initial conditions for individual trajectories using 6D hyperspherical coordinates. First, we make use of a canonical transformation to bring our system of three bodies in 3D space into 6D space ($3N-3$ translational degrees of freedom with $N=3$) and in 6D space, the trajectories can be viewed as the scattering of one particle of mass $\mu =  \sqrt{\frac{M_{\text{mol}}m_3m_4}{M}}$. The canonical transformation that accomplishes this is given by~\cite{Perez-Rios2020}: 

\begin{equation}
    \begin{split}
        \bm{\rho}_{\text{6D}} \ = \ \begin{pmatrix} \sqrt{\frac{\mu_{34}}{\mu}} \bm{\rho}_2 \\ \sqrt{\frac{\mu_{34,\text{mol}}}{\mu}} \bm{\rho}_3
        \end{pmatrix}
        \\ \mathbf{P}_{\text{6D}} \ = \ \begin{pmatrix}
            \sqrt{\frac{\mu}{\mu_{34}}} \mathbf{P}_2 \\ \sqrt{\frac{\mu}{\mu_{34,\text{mol}}}} \mathbf{P}_3
        \end{pmatrix}
    \end{split}
\end{equation}

which has the desirable properties that $\mathbf{P}_{\text{6D}} \ = \ \mu \bm{\dot \rho}_{\text{6D}}$ and that the kinetic energy is simply $T  =  \frac{1}{2}\mu \bm{\dot \rho}_{\text{6D}}^2$. The Hamiltonian under this transformation reads: 

\begin{equation}
    H \ = \ \frac{\mathbf{J}^2}{2\mu_{12}r_e^2} \ + \ \frac{\mathbf{P}^2_{\text{6D}}}{2\mu} \ + \ V(\mathbf{\hat r},\bm{\rho}_{\text{6D}})
\end{equation}
    
 The initialization of the trajectories in hyperspherical coordinates follows the procedure outlined in Ref.~\onlinecite{Perez-Rios2020}. Before sampling initial conditions for individual trajectories, two quantities must be specified: the collision energy $E_c$ and the magnitude of the impact parameter vector $b$. The magnitude of the initial 6D momentum is given by $|\mathbf{P}_{\text{6D},0}| \ = \ \sqrt{2\mu E_c}$. The direction of $\mathbf{P}_{\text{6D},0}$ is sampled using 6D hyperspherical coordinates:

 \begin{equation}
     \mathbf{P}_{\text{6D},0} \ = \ |\mathbf{P}_{\text{6D},0}|\begin{pmatrix}
        \text{sin}\alpha_{1,P} \text{sin}\alpha_{2,P} \text{sin}\alpha_{5,P} \\
        \text{cos}\alpha_{1,P} \text{sin}\alpha_{2,P}\text{sin}\alpha_{5,P} \\
        \text{cos}\alpha_{2,P}\text{sin}\alpha_{5,P} \\ 0 \\ 0 \\
        \text{cos}\alpha_{5,P}
    \end{pmatrix}
 \end{equation}
where the $\alpha_{i,P}, \  i=1,2, 5$ are the hyperangles of the initial 6D momentum. Without loss of generality, the $z$-axis in ordinary 3D space is chosen to align with the initial direction of $\mathbf{P}_3$. In general, hyperangles are sampled using the appropriate solid angle element $\Pi_{i=1}^5 \text{sin}^{i-1}\alpha_{i}d\alpha_i$ with $\alpha_1 \in [0, 2\pi)$ and $\alpha_{i>1} \in [0, \pi]$.

The impact parameter vector $\mathbf{b}$ is also sampled using hyperspherical coordinates:

\begin{equation}
    \mathbf{b} \ = \ b \begin{pmatrix} \text{sin}\alpha_{1,b}\text{sin}\alpha_{2,b}\text{sin}\alpha_{3,b}\text{sin}\alpha_{4,b} \\
    \text{cos}\alpha_{1,b} \text{sin}\alpha_{2,b}\text{sin}\alpha_{3,b}\text{sin}\alpha_{4,b} \\
    \text{cos}\alpha_{2,b} \text{sin}\alpha_{3,b} \text{sin}\alpha_{4,b} \\ \text{cos}\alpha_{3,b} \text{sin}\alpha_{4,b} \\ \text{cos}\alpha_{4,b} \\ 0 
        
    \end{pmatrix}
\end{equation}
where the basis used for $\mathbf{b}$ is forced, using the Gram--Schmidt process, to lie entirely in the 5D hyperplane perpendicular to $\mathbf{P}_{\text{6D},0}$. With $\mathbf{P}_{\text{6D},0}$ and $\mathbf{b}$ we obtain the initial 6D position vector:
\begin{equation}
    \bm{\rho}_{\text{6D},0} \ = \ \mathbf{b} \ - \ \frac{\sqrt{\rho_0^2 - b^2}}{|\mathbf{P}_{\text{6D},0}|} \mathbf{P}_{\text{6D},0}
\end{equation}
where the initial hyperradius $\rho_0$ is uniformly sampled in a range where the absolute potential energy is expected to be $<1\%$ of $E_c$ (so that trajectories are initiated in uniform rectilinear motion).

For the rotor, we return to ordinary 3D space and sample its initial conditions, the orientations of $\mathbf{\hat r}$ and $\mathbf{J}$. The orientation of $\mathbf{\hat r}$ is sampled using: 
\begin{equation}
    \mathbf{\hat r} \ = \ \text{sin}\theta \text{cos}\phi \mathbf{e}_x \ + \ \text{sin}\theta \text{sin}\phi \mathbf{e}_y \ + \ \text{cos}\theta \mathbf{e}_z
\end{equation}
where $\theta \in [0,\pi]$ and $\phi \in [0, 2\pi)$, and random sampling uses the solid angle element $\text{sin}\theta d\theta d\phi$. The initial magnitude of the angular momentum is set by the rotational quantum number $j$: 

\begin{equation}
    |\mathbf{J}_0| \ = \ \hbar \sqrt{j(j+1)}, \ j = 0,1,2, \dots
\end{equation}
The direction of $\mathbf{J}_0$ is randomly sampled in the plane perpendicular to $\mathbf{\hat r}$: 

\begin{equation}
    \begin{split}
       \mathbf{J}_0  =  |\mathbf{J}_0| \begin{pmatrix}
            \text{sin}\phi\text{cos}\eta  +  \text{cos}\theta \text{cos}\phi \text{sin}\eta \\ -\text{cos}\phi \text{cos}\eta  +  \text{cos}\theta \text{sin}\phi \text{sin}\eta \\ -\text{sin}\theta \text{sin}\eta
        \end{pmatrix}, \\ \eta \sim \text{Uniform}(0, 2\pi). 
    \end{split}
\end{equation}
For classical trajectory simulations, care must be taken with the special case of $j=0$ when working with collision energies similar to and below the quantum mechanical rotational excitation barrier. In this regime, a quantum mechanical rotor would not be expected to be excited from $j=0$ often or at all besides when it is bound in the triatomic molecule or when interacting strongly with the atoms when the reactants are in close proximity. At large separations where the potential is sufficiently weak, rotational excitations are quantum mechanically prohibited in the sense that there is not strong enough interaction with the translational degrees of freedom to induce change in the rotor's alignment relative to the other reactants. This quantum picture is incompatible with a purely classical treatment of $j=0$ because classical rotations can be induced from any amount of torque, however small, and the case of zero initial rotational motion would ensure that small torques at large separations alter the alignment of the classical rotor, leading to greater attraction between the reactants. The net effect of this classical alignment over large separations is a spurious alignment that significantly increases reaction probabilities and maximum impact parameters. To account for spurious alignment while maintaining a classical approach, we treat the $j=0$ rotor as frozen in its initial orientation, which is still randomly sampled but otherwise fixed, and we do not allow it to rotate. Higher $j$ initial rotational states do not result in as severe spurious alignment effects due to their inertia: with ample initial angular momentum, the classical rotor dynamics is less influenced in the long range prior to entering the region of strong interaction. Of course, at very high collision energies the effects of spurious alignment would be expected to decrease and become negligible, even for $j=0$ initialized trajectories, because the trajectories do not need to be initialized at as large separations, the reactants reach the interaction region faster, and the collisional energy would dominate the additional interaction energy provided by spurious alignment. Results for an unfrozen rotor initialized with $j=0$ are discussed in the Appendix.




\subsection{Interaction Potential}

The interaction potential in this work is modeled as a sum of pairwise interactions between the three reactants. In general, it is possible to include a three-body term in the potential, but for the low collision energies of interest in this work, we neglect the effect of a three-body term. Accordingly, the potential is given by:

\begin{equation}
    V(\mathbf{\hat r}, \bm{\rho}_2, \bm{\rho}_3)  =  V_{aa}(|\bm{\rho}_2|)  +  V_{ma}(R_1, \text{cos}\theta_1)  +  V_{ma}(R_2, \text{cos}\theta_2)
\end{equation}
where $V_{aa}(|\bm{\rho}_2|)$ is the atom-atom interaction and each $V_{ma}(R_i, \text{cos}\theta_i), \ i=1,2$ is a molecule-atom interaction. We use the HFD-B3-FCI1 He-He potential of Aziz~\cite{Aziz} for the $V_{aa}$ relevant to the reactions addressed in this work. The $R_i$ are the distances from the center-of-mass of the diatom to each atom, and the angles $\theta_i$ are the angles between $\mathbf{\hat r}$ and the displacements of the atoms relative to the diatom's center-of-mass. Generally, the molecule-atom interaction is anisotropic, and for the simple case of a rigid diatom-atom interaction we can write: 

\begin{equation}
    V_{ma}(R, \text{cos}\theta) \ = \ \sum_{l=0}^{l_{\text{max}}} V_l(R) P_l(\text{cos}\theta)
\label{V_expansion}
\end{equation}

\subsection{Trajectory Classification}

For individual trajectories, the scattering is considered complete either when sufficient time has passed relative to the characteristic time to reach the scattering center or when the hyperradius of the three-body system becomes sufficiently large relative to its initial value. Once the scattering is complete, the product is classified by the sign of the pairwise relative energies. For any two of the three reactants, the relative kinetic energy is computed using the relative velocity and the two-body reduced mass (e.g., the diatom and atom of mass $m_3$ have relative kinetic energy $\frac{1}{2} (\frac{M_{\text{mol}}m_3}{M_{\text{mol}}+m_3})(\mathbf{\dot r}_{\text{mol}} - \mathbf{\dot r}_3)^2$). The pairwise potential, relative kinetic energy, and, when present, the diatom's rotational energy, are added, and the two reactants are considered bound in a molecule if this energy is $<$0. 

In our treatment, there are three possible outcomes: (1) the diatom and either of the atoms combine to form a triatomic molecule, (2) the atoms form a diatomic molecule, and (3) the three reactants dissociate. If trajectories are not propagated for a long enough time, a possible fourth outcome is the formation of a complex with all reactants; however, since the total energy is positive, the system must eventually result in one of (1)--(3).

Separately, there is a classification that concerns the initial configuration of trajectories. If an initial configuration has an absolute potential energy too large relative to $E_c$, the trajectory is rejected and the equations of motion are not integrated. It is typical to set a small rejection threshold for the absolute value of the potential, and in this work we reject initial configurations with absolute potential $\geq 1 \%$ of $E_c$.

\subsection{Rate Constants}

Equipped with the machinery for integrating the classical equations of motion and sampling initial conditions, the cross-sections
and rates are calculated using the Monte-Carlo approach by calculating many trajectories for each of a number of selected ($E_c$, $b$) pairs. In this work, we run 10,000 trajectories for each ($E_c$, $b$) pair. For a given ($E_c, b$) the probability of a reaction (also known as the opacity function) is calculated as~\cite{Perez-Rios2020}:

\begin{equation}
    P_r(E_c, b) \ = \ \frac{N_r}{N} \ \pm \ \frac{\sqrt{N_r}}{N} \sqrt{\frac{N-N_r}{N}}
\end{equation}
where $N$ is the total number of (accepted) trajectories and $N_r$ is the number of trajectories that resulted in the reaction of interest. With the opacity function, the reaction cross-section, which is the effective 5D cross-sectional volume for reactions, is calculated using~\cite{Perez-Rios2020}:

\begin{equation}
    \sigma_r(E_c) \ = \ \frac{8\pi^2}{3}\int_0^{b_{\text{max}}} P_r(E_c, b)b^4 db
\label{cross_sec}
\end{equation}
where $b_{\text{max}}$ is the maximum impact parameter defined by $ P_r(E_c, b>b_{\text{max}})\approx 0$. 

Since the integral in Eq.~(\ref{cross_sec}) has a weight of $b^4$, special care must be taken when selecting the value of $b_{\text{max}}$ from Monte-Carlo calculated results. For large $b$ near $b_{\text{max}}$, there may be some small jitter in the calculated probabilities, often resulting from only one or a few reactive trajectories out of a finite number of $N$ total trajectories. As such, a method for selecting $b_{\text{max}}$ must be chosen that captures the physically relevant range of impact parameters while not counting exceedingly rare events that occur at large $b$-values that would be disproportionately weighted due to finite sampling in the Monte-Carlo approach. For example, $b_{\text{max}}$ can be selected based on how many consecutive $b$-values have probability below a chosen threshold~\cite{Py3BR}, or it may be chosen by fitting, when possible, an exponential curve to the decaying region of the opacity function and using the characteristic length scale of decay to select the location of $b_{\text{max}}$. In this work, a combination of these two methods is used to select $b_{\text{max}}$ values as consistently as possible.

Using the initial speed of the 6D particle of mass $\mu$, the volume of formation for the three-body recombination reaction is given as the energy-dependent rate constant: 
\begin{equation}
    k_3(E_c) \ = \ \sqrt{\frac{2E_c}{\mu}} \sigma_r(E_c)
\label{k3_Ec}
\end{equation}
From Eq.~(\ref{k3_Ec}), the temperature-dependent rate constant is obtained upon thermal averaging~\cite{Perez-Rios2020}:

\begin{equation}
    k_3(T) \ = \ \frac{1}{2(k_BT)^3} \int_0^\infty E_c^2 k_3(E_c)e^{-E_c/(k_BT)} dE_c
\label{k3_T}
\end{equation}

\section{Results}

\subsection{Potential Energy Surface Calculation}

The potential energy surface (PES) for the CH$^+$-He interaction is parameterized by the separation $R$ of the He atom from the center-of-mass of CH$^+$ and the angle $\theta$ between the separation vector and the CH$^+$ bond, with $\theta = 0$ corresponding to placing the He atom near H. It should be noted that the C-H separation was fixed at the equilibrium separation value of CH$^+$ of 1.1308843~\AA~\cite{CHp_re}. The energies for this closed-shell system were calculated at the CCSD(T) level of theory with counterpoise correction and basis aug-cc-pVQZ~\cite{basis1, basis2, basis3} using the Psi4 quantum chemistry package~\cite{psi4}. 

\begin{figure}
    \centering
    \includegraphics[width=1.0\linewidth]{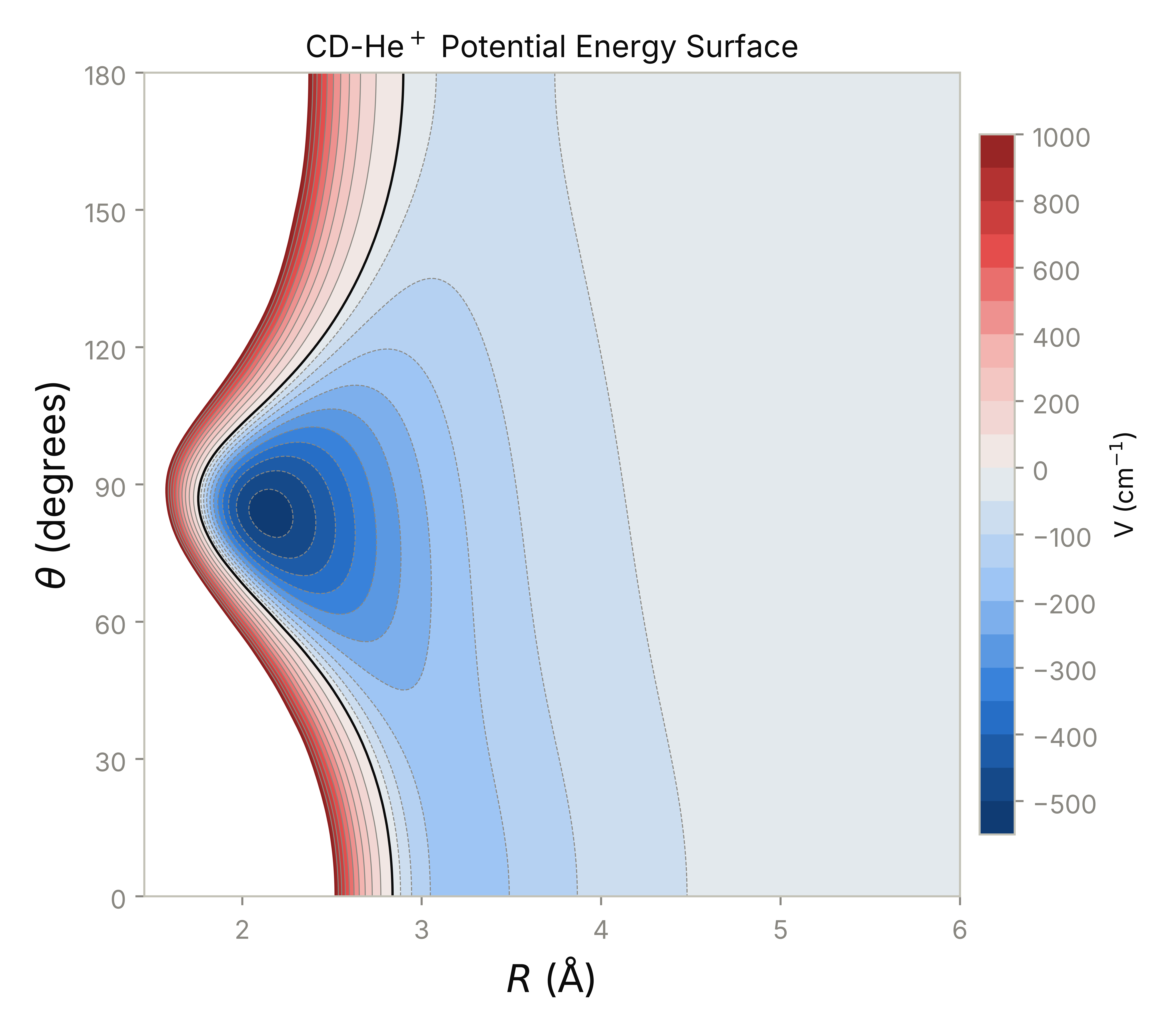}
    \caption{A contour plot of the PES calculated for CH$^+$-He for $R \in [1,6]$~\AA.}
    \label{contour_CDHe+}
\end{figure}

Excluding the calculated points with energies $>1,000$~cm$^{-1}$, which are irrelevant for the collisional energies used in this work, 1,041 points are included for the PES covering an $R$ range of [1.711, 50.0]~\AA\ and the full [0,$\pi$] range for $\theta$, which is uniformly sampled for a total of 20 $\theta$ values. A representative contour plot of the attractive well of the PES is provided in Fig.~\ref{contour_CDHe+}. In agreement with the calculation of Meuwly and Wright, we obtain a T-shaped minimum energy configuration~\cite{CHHe+_PES}. We obtain an interpolated well depth of ${\sim}525$~cm$^{-1}$, and we also report a nearby, directly calculated point of ($R , \theta, V$) = ($2.1684$~\AA, $\ang{85.26}$, $-522$~cm$^{-1}$). Our calculated well depth is similar to, but deeper than, Meuwly and Wright's value of $-477$~cm$^{-1}$~\cite{CHHe+_PES}. It should be noted that the basis used in that work is aug-cc-pVTZ, different from our extended quadruple basis. 

The CH$^+$-He PES is fit with an expansion of the form Eq.~(\ref{V_expansion}). Additionally, the functions $V_l(R)$ are taken as sums of inverse powers of $R$ (i.e., $\sum_i \frac{C_i}{R^i}$). The isotropic channel $l=0$ is equipped with the expected charge-induced dipole asymptotic interaction $-\frac{\alpha}{2R^4}$ where $\alpha$ is the polarizability of He in a.u., and no other channel has an $R^{-4}$ term. The resulting fit with $l_{\text{max}} = 7$ has an RMSE of 0.5~cm$^{-1}$ and maximum absolute deviation of 3.1~cm$^{-1}$ (41 tunable parameters, 1,041 data points). 

\subsection{Three-Body Recombination Rates}

\begin{figure}
    \centering
    \includegraphics[width=1.0\linewidth]{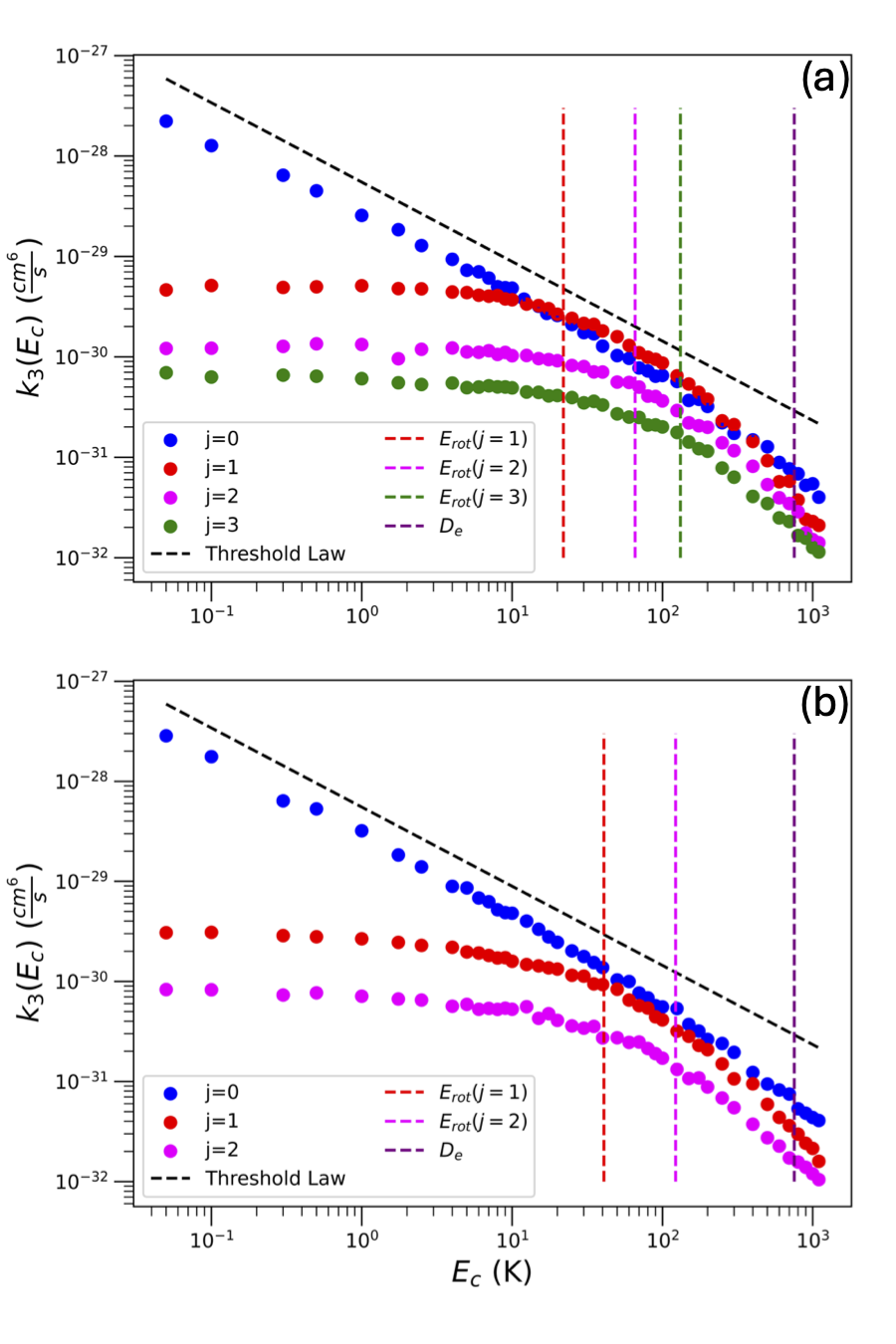}
    \caption{The collision energy-dependent three-body recombination rate constants $k_3(E_c)$ for the reactions studied in this work. Panel (a) corresponds to CD$^+$ + He + He $\rightarrow$ CDHe$^+$ + He and panel (b) corresponds to the same attachment reaction but with CD$^+$ replaced by CH$^+$. The black-dashed lines give the rates as calculated by the threshold law for ion-atom-atom three-body recombination. The rotational energy levels of the rotor are also marked.  }
    \label{Ec_rates}
\end{figure}

Using the methods presented in Sec.~II, the attachment reactions CD$^+$ + He + He $\rightarrow$ CDHe$^+$ + He and CH$^+$ + He + He $\rightarrow$ CHHe$^+$ + He are studied over a wide range of collision energies and for initial rotational numbers $j = 0,1,2$ (and additionally $j=3$ for the reaction with CD$^+$). While He$_2$ formation is possible, its probability is low compared to CDHe$^+$/CHHe$^+$ formation, as expected in light of the stronger charge-induced dipole interaction versus the neutral-neutral interaction~\cite{Mirahmadi_2023}. The results for the energy-dependent three-body recombination rate constants $k_3(E_c)$ for CD$^+$ + He + He $\rightarrow$ CDHe$^+$ + He and CH$^+$ + He + He $\rightarrow$ CHHe$^+$ + He are shown in panels (a) and (b) of Fig.~\ref{Ec_rates}, respectively. As discussed above, the results for $j$=0 were obtained by freezing the diatom's rotation, so during the course of the collision the molecular ion does not rotate and maintains its original orientation. This is necessary to avoid the spurious alignment that will occur in a classical scenario when the molecule is free to rotate without initial angular momentum. In addition, we note that for collision energies below 1~K only 3--8 partial waves contribute, and since our approach is completely classical, our method would not capture any possible resonance in this energy regime.


The most striking feature of Fig.~\ref{Ec_rates} is the flattening of the $k_3(E_c)$ curve in the regime $E_c \ll E_{\text{rot}}$ for $j>0$. According to the threshold law for ion-atom-atom recombination applicable at low collision energies and dictated by long-range physics, the energy-dependent three-body recombination rate constant follows a power law  $k_3(E_c) \propto E_c^{\frac{10 + \beta}{2\beta}}$, where $\beta$ appears in the approximate, hyperradius-dependent potential applicable at long-range: $V_{LR}(\rho_{\text{6D}})  =  -C_{\text{eff}}\rho_{\text{6D}}^{\beta}, \ \beta <0$ ~\cite{Mirahmadi02102022}. The threshold law is derived under the capture model assumption that if $E_c$ exceeds the three-body centrifugal barrier furnished by the grand angular momentum, then the reaction will occur with unit probability; otherwise, if the centrifugal barrier is not exceeded, no reaction occurs. Therefore, our results for $j>0$ are dictated not by the long-range, but rather by the specifics of the rotational-translational energy coupling mechanisms related with short and medium-range separations, as we explain below. The short and medium-range details of the trajectories (i.e., effects that emerge from the rotational motion influencing the reactants when either, or both, of the atoms are close to the diatom) matter and dominate the behavior of $k_3(E_c)$ when the collision energy is much lower than the initial rotational energy of the diatom. When present, the energy and motion of the internal, rotational degrees of freedom, on average, disrupt and prevent attachment, resulting in lower rates and little dependence on the translational energy $E_c$ in this regime. As the collision energy increases towards and exceeds the initial rotational energy, the $k_3(E_c)$ curves start to behave more like power laws of similar exponent. This indicates that the threshold law becomes applicable as the translational energy becomes similar to and greater than the initial rotational energy. This is not surprising: as the collision energy becomes similar to the initial rotational energy, the effects of the energy and motion of the internal degrees of freedom cease to dominate the behavior of $k_3(E_c)$ and the threshold law becomes applicable. Clearly, though, the effect of the initial rotational energy is present as we still have $k_3(E_c; j=3) <  k_3(E_c; j=2) <  k_3(E_c; j=1) $ even when the power law behavior returns. 

Another notable feature of Fig.~\ref{Ec_rates} is that, in panel (a), the $j=1$ curve intersects and exceeds the $j=0$ curve for $ E_{\text{rot}}(j=1)\lesssim E_c \lesssim 250 $~K. This behavior is absent in panel (b). The intersection may arise from spurious alignment effects similar to those that led us to freeze the rotor for $j=0$. Because the rotational constant of CD$^+$ is half that of CH$^+$, the rotor may align during the collision, thereby enhancing the reaction rate.

\begin{figure}
    \centering
    \includegraphics[width=1.0\linewidth]{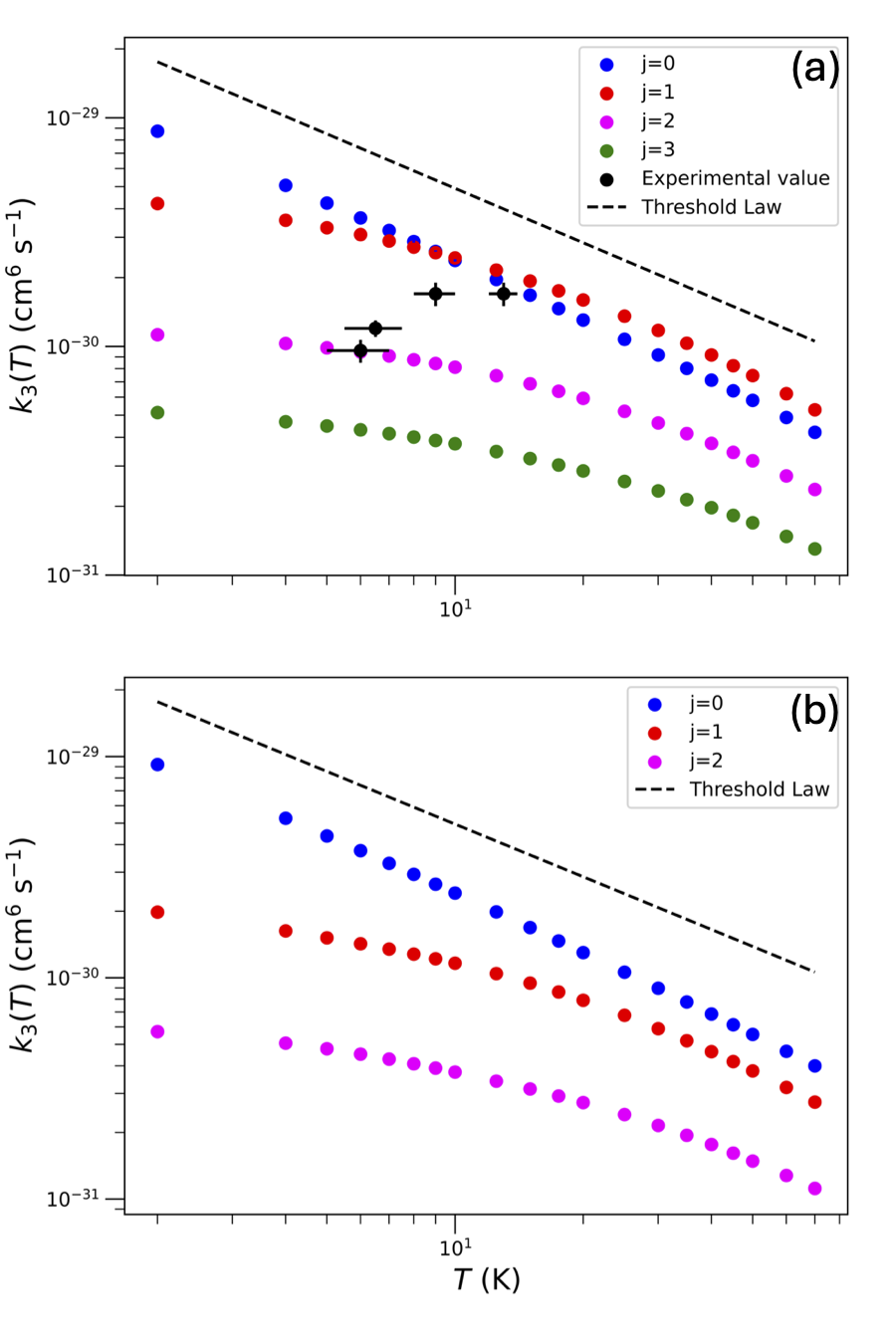}
    \caption{Thermally averaged rate constants obtained using Eq.~(\ref{k3_T}) and the data in Fig.~\ref{Ec_rates}. Panels (a)-(b) here correspond to the same reactions of panels (a)-(b) in Fig.~\ref{Ec_rates}. In panel (a), experimental data is obtained from Ref.~\onlinecite{BRUNKEN201767} by using the collisional temperature instead of the the trap temperature.}
    \label{thermal_rates}
\end{figure}

Using Eq.~(\ref{k3_T}), we calculate the thermally averaged three-body recombination rate constants for CD$^+$ + He + He $\rightarrow$ CDHe$^+$ + He and CH$^+$ + He + He $\rightarrow$ CHHe$^+$ + He, including $j=0,1,2$ at temperatures between 2--70~K, as shown in panels (a) and (b) of Fig.~\ref{thermal_rates}, respectively. In panel (a), we compare our results with the available experimental data from Ref.~\onlinecite{BRUNKEN201767}. The thermally averaged rate constants follow a trend similar to the calculated $E_c$-dependent ones: $j>0$ rate constants follow a power law for $E_c \gtrsim E_{\text{rot}}$ and begin to flatten for $E_c < E_{\text{rot}}$. This behavior is even clearer in panel (b) since CH$^+$ has twice the rotational constant of CD$^+$. As a result, depending on the temperature range being explored, the temperature dependence of the three-body recombination will change. For instance, in panel (b) we observe a noticeable change in the temperature trend in the range between 10 and 70K. Furthermore, the $j>0$ curves appear to approach a common power law shared with $j=0$ as the temperature increases. As may be expected, the $j=0$ power law agrees with the threshold law predictions, although it deviates at higher temperatures due to contributions from high-energy rates that depend on the short-range of the interaction potential.

With respect to the experimental data, we find good agreement, specifically at 10 and 14 K, although the trend of the experimental data is not reproduced by our simulations. The experimental rates increase with temperature, a trend that could arise from an activation barrier or resonance effects; however, three-body recombination is a barrierless process and therefore shows a rate inversely proportional to temperature. Subsequent experimental studies of trapped ion--neutral collisions have reported similar trends, potentially due to uncertainties in ion-temperature characterization~\cite {Wester_2021}. Revisiting the temperature characterization in these measurements could therefore help determine whether this uncertainty accounts for the discrepancy with our simulations.

\subsection{Opacity Functions}

As discussed above, the opacity function $P_r(E_c, b)$ is the probability of the reaction denoted by $r$ to occur at collision energy $E_c$ with impact parameter $b$. Computationally, it is obtained by running many trajectories whose initial conditions are randomly sampled according to Sec.~II~A. The opacity function is used for calculating the cross-section [Eq.~(\ref{cross_sec})] which is then used for determining rate constants [Eqs.~(\ref{k3_Ec}) and (\ref{k3_T})]. The opacity function is usually represented as the probability of reaction as a function of $b$ at fixed $E_c$, which is then integrated against $b$ with a geometric weight of $b^4$ [Eq.~(\ref{cross_sec})]. Opacity functions typically have some structure for $b$ values much smaller than $b_{\text{max}}$; global maxima, and sometimes local extrema, usually reside at small $b$ values. The probability of reaction is usually highest at or near $b=0$, which is analogous to a head-on collision. As $b$ approaches $b_{\text{max}}$, the opacity function decays to near zero. There may be some small jitter in the calculated opacity function for $b \gtrsim b_{\text{max}}$ due to the finite number of trajectories $N$ used in the Monte-Carlo approach.    

\begin{figure}[h!]
    \centering
    \includegraphics[width=1.0\linewidth]{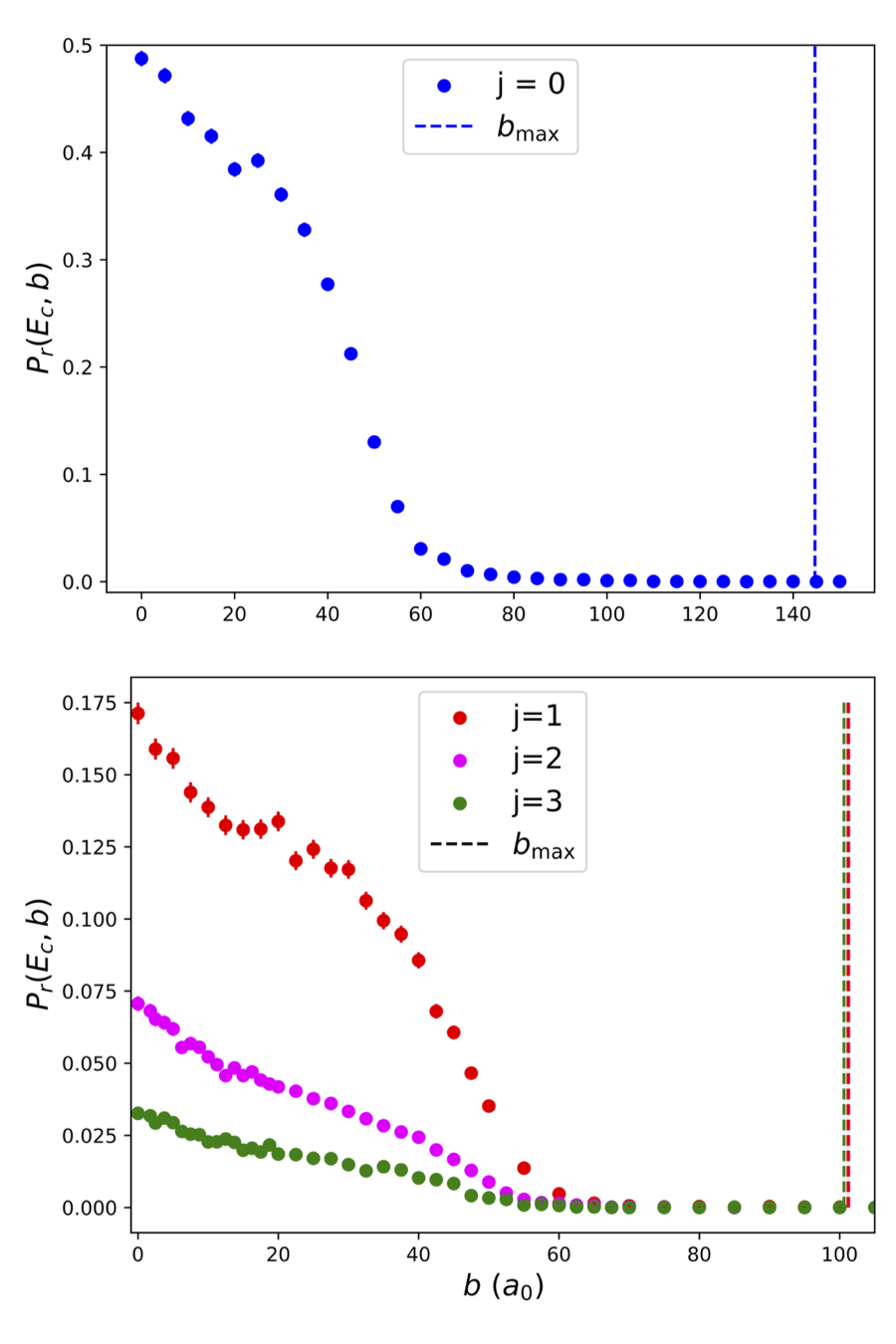}
    \caption{Opacity functions across four different values of the initial quantum number $j$ at a common collision energy $E_c = 1$~K for the reaction involving CD$^+$.}
    \label{opacity_1}
\end{figure}

Example opacity functions across initial rotational numbers $j = 0,1,2,3$ at a common collision energy of $E_c = 1$~K are given in Fig.~\ref{opacity_1}. This collision energy corresponds with the region where the $j>1$ reaction rates are flat, but the $j=0$ results follow the power law of the threshold law. First, we notice that the overall scale of the opacity for $j=0$ is much larger than the rest, which aligns with the different behaviors in the rate constants in that $j=0$ trajectories have higher reaction probabilities. In particular, the value of $b_{\text{max}}$ depends on whether $j>0$ with the $j=0$ opacity function having the largest $b_{\text{max}}$ and the $j>0$ opacity functions sharing a smaller $b_{\text{max}}$. Therefore, the long-range effects on the probability of reaction are the same for $j>0$, but distinct from $j=0$. Similarly, there is a drastic suppression of the opacity at small impact parameters for $j>0$ relative to $j=0$, emphasizing the relevance of the rotational energy of the molecular ion in the recombination process. 

\begin{figure}[h!]
    \centering
    \includegraphics[width=1.0\linewidth]{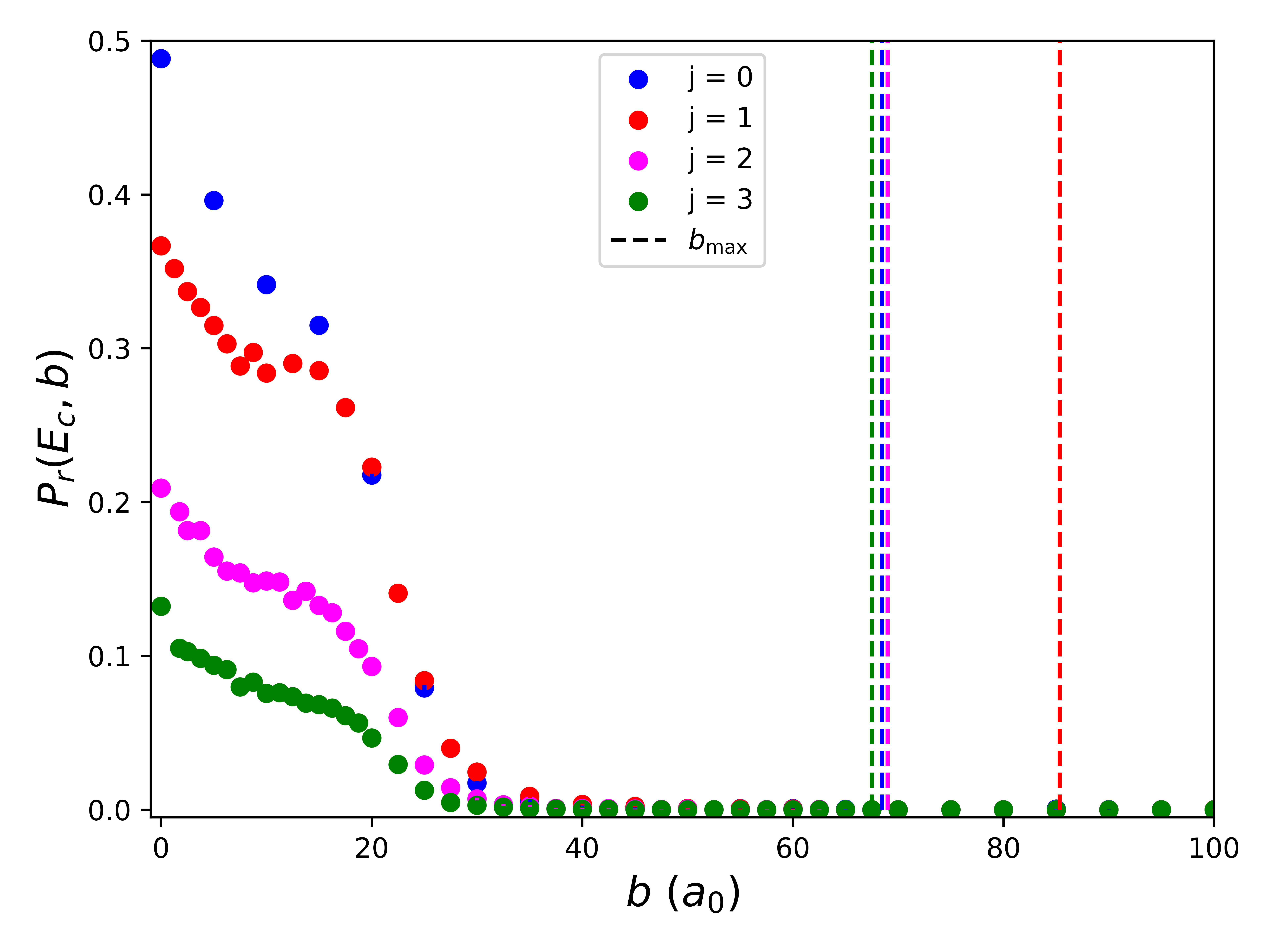}
    \caption{Opacity functions across four different values of the initial quantum number $j$ at a common collision energy $E_c = 20$~K for the reaction involving CD$^+$.}
    \label{opacity_2}
\end{figure}

Figure~\ref{opacity_2} shows the corresponding opacity functions at $E_c=20$~K. At this energy, the rate constants for $j>0$ begin to approach the power-law regime, and the opacity functions become more similar than at $E_c=1$~K. This reduced sensitivity to the initial rotational state is consistent with the increasing importance of the translational collision energy relative to the molecule's rotational energy. Except for $j=1$, the opacity functions have similar maximum impact parameters. Maximum probabilities decrease with increasing $j$, although the suppression is less pronounced than at $E_c=1$~K. Thus, apart from the $j=1$ case, the rotational excitation of initial states primarily reduces the reaction probability while leaving $b_{\mathrm{max}}$ approximately unchanged. This behavior is consistent with Fig.~\ref{Ec_rates}, where the diatom with initial rotational energy generally lowers the magnitude of the recombination rate while approximately preserving its power-law energy dependence.

The $j=1$ opacity function in Fig.~\ref{opacity_2} extends to a substantially larger impact parameter than those of the other initial rotational states, indicating that recombination remains possible in more distant encounters. One possible explanation is that coupling between rotational and orbital motion favors capture for this initial state. For higher rotational states, faster rotation may reduce the effectiveness of orientational steering by the anisotropic molecule--atom interaction. However, the opacity functions alone do not establish this mechanism. Spurious alignment of the rotor orientation incited outside of the interaction region could also contribute to the enhanced recombination probability and should be assessed before attributing the $j=1$ behavior to an effect that could be physically realized.

\subsection{Trajectories and Rotational Effects}

\begin{figure*}[t]
    \centering
    \includegraphics[width=1.0\textwidth]{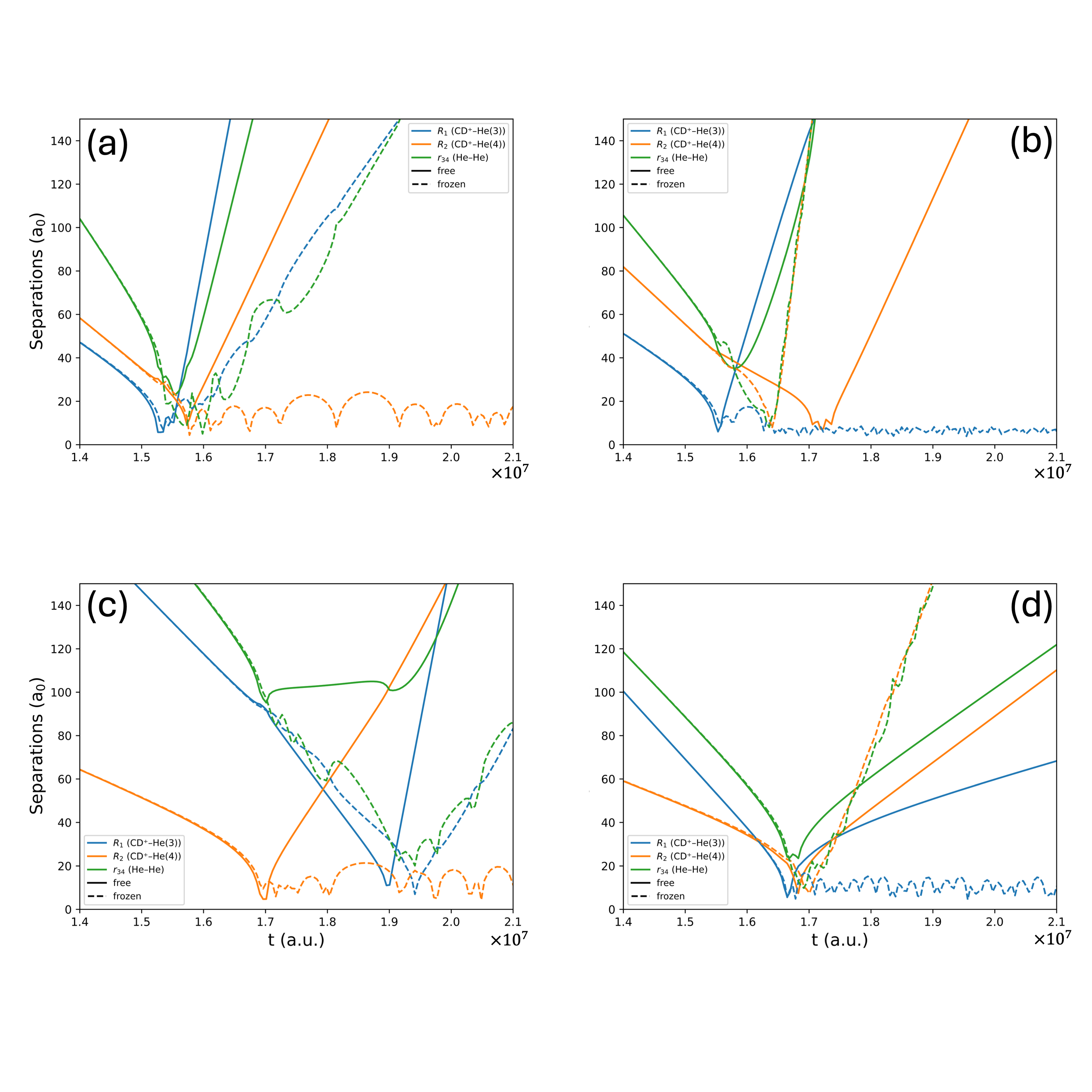}
    \caption{Example single trajectories for the CD$^+$ reaction at a collision energy of $E_c = 1$~K with panels a, b, c, and d having impact parameters $b = [0,10,20,30] \ a_0$, respectively. Each panel represents a unique set of initial conditions and shows the propagated trajectories for: the rotor free to rotate with initial quantum number $j=1$ (solid lines) and the rotor frozen in its initial orientation, unable to rotate (dashed lines) corresponding to an initial state of $j=0$. $R_1$ and $R_2$ are the separations between the center-of-mass of CD$^+$ and the He atoms 3 and 4, respectively, and $r_{34}$ is the separation between the He atoms. }
    \label{trajs}
\end{figure*}

To understand more directly why the reaction probabilities, and thus the rate constants, decrease for increasing $j$ in the $E_c \ll E_{\text{rot}}$ regime, it is instructive to look at how individual trajectories are altered by the introduction of the diatom's orientational degrees of freedom with initial rotational energy. In Fig.~\ref{trajs}, the trajectories resulting from four unique sets of initial conditions with $E_c = 1$~K and impact parameters $b = [0,10,20,30] \ a_0$, corresponding to panels (a)-(d), respectively, are displayed, showing the results for both a $j=0$ frozen rotor (dashed lines) and a rotor that is free to rotate with an initial angular momentum corresponding to $j=1$ (solid lines). The $j=0$ case is treated as a frozen rotor, having fixed orientation, for the reason of spurious alignment discussed above. The trajectories shown in Fig.~\ref{trajs} highlight cases where the $j=0$ initial state leads to the formation of the triatomic molecule CDHe$^+$, but the $j=1$ initial state results in dissociation. Each comparison is done with the same set of initial conditions. In agreement with the rate constants in Fig.~\ref{Ec_rates} at $E_c = 1$~K, it is much more common for $j=0$ trajectories to result in reactions and $j=1$ trajectories to not result in reactions than the inverse. Thus, the trajectories in Fig.~\ref{trajs} serve to give us a glimpse into possible mechanisms for how the diatom's rotational motion reduces reaction probabilities. It must be said, though, that results that come from summing over many trajectories, such as opacity functions and rate constants, represent the effect that the rotational motion has on average on the reaction, but here we will focus on individual trajectories. Also, the results presented above do not reuse the same initial conditions across different $j$-values (to ensure we do not have correlated results) but initial conditions are reused in Fig.~\ref{trajs} to isolate the effects of rotational motion.

In general, the effects of the rotational motion on the occurrence of reactions must appear at short to medium-range separations (where the term ``medium-range'' here denotes separations where the anisotropy of the molecule-atom interaction may be felt without probing the short-range repulsive walls). We would not expect the effects of the rotational motion to appear in the long-range because the anisotropy in the molecule-atom potential is weak there (this is especially the case in our system, which is dominated by the isotropic charge-induced dipole interaction in the long-range). Looking to the trajectories in Fig.~\ref{trajs}, it is indeed the case that the frozen ($j=0$) vs.\ free (with $j=1$) scenarios do not diverge until medium-range separations, far away from the repulsive wall but close enough for the anisotropy in the molecule-atom interaction to matter. 

In Fig.~\ref{trajs}, several different mechanisms can be identified that prevent the formation of the triatom in the $j=1$ scenario. In panel a, the effect of the $j=1$ rotational motion is to donate kinetic energy to the He atom labeled He(3) (not to be confused with the isotope $^3$He) ejecting it from the interaction region. In the frozen $j=0$ scenario, He(3) serves as the third body that takes away the excess kinetic energy that allows for the formation of the triatomic ion. Thus, in panel (a), the rotational motion of the diatom prevents the attachment by ejecting the third body, which is otherwise essential to the barrierless formation of the triatom. Thus, a mechanism of reaction prevention is the ejection of the third body by the spinning rotor. Of course, such an effect is going to be most important in the energy regime $E_c \ll E_{\text{rot}}$ where energy donated from rotations can dominate the incident translational energy. 

In Fig.~\ref{trajs} panel (b), the attachment is prevented in a different manner compared to panel (a). Here, the He atom labeled He(3) attaches to CD$^+$ for $j=0$, but in the $j=1$ scenario, He(3) is ejected when it gets close to the molecule before the other He atom arrives. In the case of $j=0$, He(3) is allowed to attach to the molecule for a brief period of time (and make two bounces) before the other He atom arrives to take away the excess kinetic energy, which is needed to complete the attachment. The overall effect of the rotational motion here is to prevent the attachment of the would-be-bonding He atom to the diatom and subsequently ejecting it. Looking more closely, He(3) actually gets brought closer to the diatom sooner in the $j=1$ scenario than in the $j=0$ scenario, indicating that the rotational motion had the effect of bringing He(3) in with more force to where it likely felt the repulsive wall of the molecule-atom potential, ultimately leading to its ejection. Generally, it is plausible that the excess rotational energy and motion present for $j>0$ can prevent the initial attachment that occurs before the third body arrives and removes excess kinetic energy.

In Fig.~\ref{trajs} panel (c), we have another example of the rotational motion prohibiting the initial attachment of the would-be-bonding He atom. Like in panel b, the rotating diatom attracts a He atom (this time He(4)) closer than it would in the $j=0$ scenario, after which it is ejected. What is notable about panel (c) is how the prohibited attachment occurs when the third body, which here is He(3), is quite distant from the other reactants. This demonstrates that the prohibited initial attachment, resulting from the diatom's rotational motion, happens in a two-body context.

In Fig.~\ref{trajs} panel (d), we see an example of the diatom's rotational motion preventing the reaction in a three-body context. In the $j=1$ scenario, both atoms approach nearly equally close to the diatom before being ejected from the interaction region. In the $j=1$ scenario, the third body (He(4)) is brought close to the diatom sooner than in the $j=0$ scenario. In the $j=0$ scenario, the diatom-atom complex containing He(3) is allowed to bounce almost twice before the third body arrives to take away excess kinetic energy, whereas a smaller time delay between the arrival of the third body is present in the $j=1$ scenario. For $j=1$, the rotational motion prevents attachment seeing as He(3) begins to be ejected before the closest approach of He(4) to the diatom.

The mechanisms that suppress reactions discussed in relation to Fig.~\ref{trajs} occur at short to medium range separations and are explained by the presence of excess rotational energy and the fact that the diatom is rotating. The effects of the rotations of the diatom are most important when $E_c \ll E_{\text{rot}}$. Since the diatom's initial rotational energy dominates in this regime, it is sensible that the $j>0$ curves flatten in Fig.~\ref{Ec_rates}, showing no drastic changes with varying $E_c$. Without internal degrees of freedom playing a role, we would anticipate $k_3(E_c)$ to follow a power law for $E_c$ far below the dissociation energy threshold. This power law emerges under the assumption of capture theory; in our case, that would mean that the dominant reaction occurs with unit probability if the long-range centrifugal barrier associated with the grand angular momentum is overcome by $E_c$, allowing the reactants to be captured. Of course, this neglects any short or medium range effects that may occur below the long-range centrifugal barrier. In our case with excited internal degrees of freedom, the capture assumption does not hold and interesting effects do arise at medium and short separations. As the collision energy approaches and exceeds the diatom's initial rotational energy, the power law appears since the translational energy is no longer dominated by the rotational energy. Even when the power law reappears, the magnitude of the rate constants decrease with increasing $j$, so mechanisms of reaction suppression are present but not dominant in this regime.

\section{Conclusions}

We have introduced a classical trajectory approach in hyperspherical coordinates for studying diatom--atom--atom three-body recombination, explicitly including the rotational degrees of freedom of the diatomic molecule. The relative translational motion is represented in a six-dimensional configuration space described by hyperspherical coordinates, while the initial rotational conditions are chosen to reproduce the energy of the corresponding quantum rotational state.

Using CD$^+$/CH$^+$ + He + He $\rightarrow$ CDHe$^+$/CHHe$^+$ + He as a proof of concept, we find that the classical threshold scaling established for atomic three-body recombination is recovered when the collision energy exceeds the initial rotational energy of the molecular ion. Below this rotational energy scale, the recombination rate constants become approximately independent of collision energy. Our results indicate that this behavior arises from coupling between molecular rotation and the relative translational motion of the three collision partners. These findings suggest that molecular ions with larger rotational constants should exhibit this regime at higher collision energies. We therefore propose H$_2^+$ + He + He $\rightarrow$ H$_2$He$^+$ + He as a candidate for investigating this effect at temperatures around 90~K, compared with approximately 20-40~K for the system studied here.

Our calculated thermal rate constants agree well with the available experimental values at 10 and 14~K, although they do not reproduce the observed temperature dependence. The experimental rates increase with temperature and appear to approach a plateau, in contrast to the decreasing dependence predicted by our long-range capture description of barrierless recombination. This discrepancy could reflect additional dynamical effects, such as resonances, or uncertainties in the characterization of the experimental collision energies. However, similar trends have been shown in other experiments, indicating that a recalibration of the temperature is required for this system, and a new experimental rate will be very useful to establish the accuracy of the present approach.

The approach can also be extended to other diatom--atom--atom reactions, including O$_2$ + O + Ar and S$_2$ + S + Ar. It thus provides a framework for exploring the role of molecular rotation in a broader range of three-body association processes involving four constituent atoms.

\begin{acknowledgments}
This work was supported by the U.S. National Science Foundation through CAREER Award No. PHY-2440808. We appreciate the insights of Pavol Jusko in discussions about how this work relates to potential experiments.
\end{acknowledgments}

\section*{Author Declarations}
\subsection*{Conflict of Interest}
The authors have no conflicts of interest to declare.
\subsection*{Author Contributions}
\textbf{Daniel Julian:} data curation, formal analysis, methodology, visualization, writing-original draft.  \textbf{Jes\'us P\'erez-R\'ios:} conceptualization, formal analysis, funding acquisition, methodology, writing-original draft.
\section*{Data Availability}
The data that support the findings of this work are available from the corresponding author upon reasonable request.

\appendix*

\section{$j=0$ Unfrozen Rotor Rate Constants}

\begin{figure}[h!]
    \centering
    \includegraphics[width=1.0\linewidth]{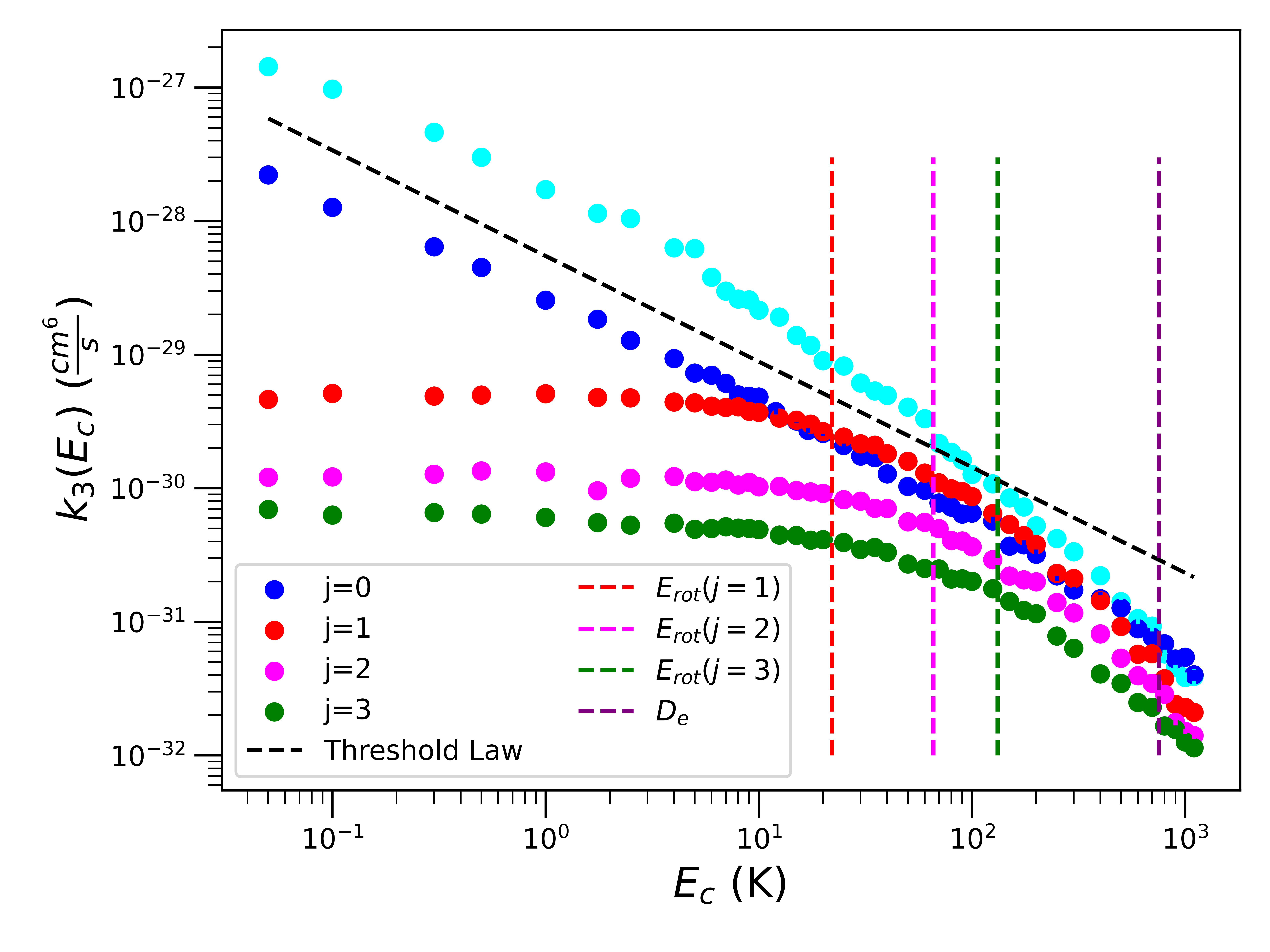}
    \caption{Displayed above are the results shown in Fig.~\ref{Ec_rates} panel (a) with an additional set of points (aqua blue) for the energy-dependent rate constants obtained for $j=0$ without freezing the rotor; in other words, trajectories run to obtain these points allow the rotor to freely rotate with zero initial angular momentum.}
    \label{unfrozen_j0}
\end{figure}

In Fig.~\ref{unfrozen_j0}, $k_3(E_c)$ for CD$^+$ + He + He $\rightarrow$ CDHe$^+$ + He was also calculated for initial states corresponding to $j=0$ of CD$^+$, but without freezing the rotational dynamics of the diatom (the aqua blue data points). Without freezing the rotor, the classical, rotational dynamics are calculated, so spurious alignment of the diatom is allowed during the approach of the diatom in the long-range. Unfreezing the rotor with an initial rotational state corresponding to $j=0$ results in rate constants almost an order of magnitude too large compared to the frozen case at low collision energies. Also, the unfrozen, $j=0$ rate constants are larger than the threshold law values for the corresponding ion-atom-atom recombination (which does not include internal degrees of freedom). Importantly, the rate constants for the unfrozen, $j=0$ rotor would be too large to produce thermally averaged rate constants that agree with the experimental values (See Fig.~\ref{thermal_rates}). Unsurprisingly, at the highest collision energies considered, the unfrozen, $j=0$ results start to agree with the other calculated rate constants.

%

\end{document}